\documentclass[openacc]{rstransa}

\begin{document}

\title{Asteroid-Comet Continuum Objects in the Solar System}

\author{
Henry H.\ Hsieh$^{1,2}$}

\address{$^{1}$Planetary Science Institute, 1700 East Fort Lowell Rd., Suite 106, Tucson, Arizona 85719, USA\\
$^{2}$Institute of Astronomy and Astrophysics, Academia Sinica, P.O.\ Box 23-141, Taipei 10617, Taiwan}

\subject{Solar system}


\corres{H.\ H.\ Hsieh\\
\email{hhsieh@psi.edu}}

\begin{abstract}
In this review presented at the Royal Society meeting, ``Cometary Science After Rosetta'', I present an overview of studies of small solar system objects that exhibit properties of both asteroids and comets (with a focus on so-called active asteroids).  Sometimes referred to as ``transition objects'', these bodies are perhaps more appropriately described as ``continuum objects'', to reflect the notion that rather than necessarily representing actual transitional evolutionary states between asteroids and comets, they simply belong to the general population of small solar system bodies that happen to exhibit a continuous range of observational, physical, and dynamical properties.  Continuum objects are intriguing because they possess many of the properties that make classical comets interesting to study (e.g., relatively primitive compositions, ejection of surface and subsurface material into space where it can be more easily studied, and orbital properties that allow us to sample material from distant parts of the solar system that would otherwise be inaccessible), while allowing us to study regions of the solar system that are not sampled by classical comets.
\end{abstract}


\begin{fmtext}
\section{Background}

Asteroids are classically understood to be essentially inert objects composed primarily of non-volatile material.  They are mostly found in the inner solar system (inside the orbit of Jupiter) where they are believed to have formed.  Meanwhile, comets are classically thought of as ice-rich bodies originally from the outer solar system (beyond the orbit of Neptune, in the Kuiper Belt, scattered disk, or Oort Cloud) that have been perturbed onto orbits passing through the inner solar system.  While in the inner solar system, when they are sufficiently close to the Sun and therefore sufficiently heated, sublimation
\looseness=-1

\end{fmtext}
\maketitle

\noindent of their volatile contents drives the release of gas and dust, producing cometary activity in the form of comae, tails, or both.

In much of comet research, which typically focuses on ``classical'' Jupiter-family comets (JFCs), Halley-type comets (HTCs), and long-period comets (LPCs), the underlying fundamental goal is to infer details about the temperature, compositional, and dynamical structure of the early solar system, and also to learn about the solar system's formation and evolution.  This is also a frequent objective of studies of asteroids and meteorites, but comets have certain characteristics that make them particularly valuable for addressing these topics.
\looseness=-1

First, the present-day icy states of comets mean that their contents are likely to be more primitive and represent better preserved samples of the early solar system than asteroid or meteorite material.  In the cases of asteroids and meteorites, the lack of substantial extant ice (and in many cases, the presence of hydrated minerals that formed in the presence of liquid water) points to a history of significant thermal processing through radiogenic or solar heating, while still-frozen cometary ice is likely to have undergone at least somewhat less thermal processing (although is probably also not completely pristine).

Second, the very nature of cometary activity means that material from the surface, and sometimes the sub-surface, of the cometary object is being launched into space.  There, it becomes subject to an array of remote and in situ analysis methods that cannot be applied to the inert solid surface of an inactive asteroid (e.g., see other Rosetta-related articles in this issue).

Lastly, classical comets typically have orbits that bring them from trans-Neptunian space to close proximity to the Earth, where they can be studied in far greater detail than they could in their (assumed) original source regions in the outer solar system.  As they move along their highly eccentric orbits, comets also undergo dramatic changes in temperature which can provide insights into their compositions given the different temperatures at which various volatile species sublimate.  These dynamical properties therefore essentially allow us to study samples of the outer solar system at a level of detail and over a range of environmental conditions that would otherwise be impossible with our current technological capabilities.

The number of objects exhibiting observational, physical, and dynamical properties of both comets and asteroids has grown steadily in recent years as surveys have turned up examples of rare objects and physical analyses have become more sophisticated.  As such, the traditional distinctions between asteroids and comets have blurred, requiring us to be more specific about the exact meanings of these terms in particular contexts (recognizing that these meanings may be different in different contexts).  On the other hand, the proliferation of objects exhibiting at least some properties of comets, if not others, means that we can take advantage of some of the properties discussed above that make classical comets interesting to study in order to study other types of solar system objects using a broader range of different techniques than in the past.

In this review presented at the Royal Society meeting, ``Cometary Science After Rosetta'', I present a broad overview of studies of small solar system objects that exhibit properties of both asteroids and comets (with a particular focus on so-called active asteroids).  Sometimes referred to as ``transition objects'', these bodies are perhaps more appropriately described as ``continuum objects''.  In other words, rather than necessarily representing transitional evolutionary states via which comets evolve into asteroids or vice versa, these objects simply belong to the general population of small solar system bodies that happen to exhibit a continuous range of observational, physical, and dynamical properties of both asteroids and comets.  For additional perspectives, the reader is also referred to previous reviews in the literature \cite{ref-weissman02,ref-toth06c}.

\section{Types of Continuum Objects\label{section:continuumobjs}}

\subsection{Dormant and Extinct Comets\label{section:acos}}

Astronomers have long recognized the potential for inactive comets to be mistaken for asteroids.  By 1970, several authors had already noted the dynamical similarities between apparently inactive asteroids, such as the Apollo and Amor asteroids, with short-period comets, as well as the potential for previously known comets to appear completely inactive over certain portions of their orbits \cite{ref-kresak67,ref-marsden70}.  A hypothesis was put forth that these objects, as well as low-activity comets like 28P/Neujmin 1 and 49P/Arend-Rigaux, could represent transitional phases between comets and minor planets (i.e., asteroids).  It was noted, though, that the cometary origins of such objects could still be identified via dynamical criteria, such as their dynamical lifetimes \cite{ref-marsden70}. A more clearly defined dynamical criterion for distinguishing the orbits of asteroids and comets was eventually adopted in the form of the Tisserand invariant ($T_J$), or Tisserand parameter, which is a mostly conserved quantity in the restricted circular three-body problem, and is given by
\begin{equation}
T_J = {a_J\over a_{\rm obj}} + 2\cos(i_{\rm obj}) \left[{a_{\rm obj}\over a_J} \left(1-e_{\rm obj}^2\right)\right]^{1/2}
\end{equation}
where $a_J=5.2$~AU is the semimajor axis of Jupiter, and $a_{\rm obj}$, $e_{\rm obj}$, and $i_{\rm obj}$ are the semimajor axis, eccentricity, and inclination of the object, respectively.  Objects with $T_J<3$ are essentially dynamically coupled with Jupiter and are commonly considered to have ``comet-like'' orbits, while objects with $T_J>3$ do not have close encounters with Jupiter and are commonly considered to have ``asteroid-like'' orbits \cite{ref-kresak67,ref-vaghi73,ref-kresak80}.  Other more sophisticated sets of dynamical classification criteria have been devised and tested \cite{ref-tancredi14,ref-licandro16}, but it is probably fair to say that $T_J$ remains the most commonly used criterion for dynamically classifying objects as ``comet-like'' and ``asteroid-like'', due to its simplicity as well as its long history of use for this purpose.

Numerous researchers over the last several decades have considered the cases of comets that do not show activity at certain times, such as while far from the Sun where temperatures are too low to drive sublimation, as a result of mantling, or at the end of their active lifetimes when repeated sublimation events have depleted their volatile supplies \cite{ref-hartmann87,ref-kresak87,ref-weissman89}. 
These so-called asteroids on cometary orbits (i.e., orbits with $T_J<3$), or ACOs, have been the subjects of various observational studies aimed at confirming the cometary nature of these objects \cite{ref-hicks00,ref-carvano08} and estimating the contribution of these objects to the near-Earth object (NEO) population.  An albedo survey of 32 asteroids with a range of orbit types found a striking correlation between low $T_J$ values and albedos with 64\%$\pm$5\% of objects with $T_J<3$ having low, ``comet-like'' albedos (i.e., $p<0.075$), leading the authors to conclude that dormant or extinct comets could comprise 4\% of the known NEO population \cite{ref-fernandez05}.  This result was supported by a spectroscopic and albedo study of 55 NEOs that found that 54\%$\pm$10\% of NEOs with $T_J<3$ have ``comet-like'' spectra (i.e., consistent with C-, P-, T-, or D-type asteroids) or albedos, concluding that 8\%$\pm$5\% of NEOs could be dormant or extinct comets \cite{ref-demeo08}, where both observational studies are roughly consistent with a dynamical study that estimated a total contribution of $\sim$6\% to the current NEO population from dormant or extinct JFCs \cite{ref-bottke02}.  Other observational studies have also generally attempted to probe what may be the end state of cometary physical evolution \cite{ref-lamy09}, with some attempts to also connect these objects with the delivery of potentially cometary meteorites to the Earth \cite{ref-campins98}.

The conclusions of these studies are inevitably uncertain given the similarity between the spectra of cometary nuclei and those of the aforementioned C-, P-, T-, and D-type asteroids \cite{ref-hicks00,ref-demeo08}, which comprise most of the observed asteroids in the outer main belt, Hilda, and Jovian Trojan populations \cite{ref-gradie82,ref-dahlgren95,ref-dahlgren97,ref-grav12,ref-demeo13}, any of which could also be the source of the observed low-albedo, spectroscopically ``comet-like'' NEOs.  Identification of formerly active comets via dynamical means is also complicated by the fact that $T_J$ is actually not always conserved [Section~\ref{section:dynamics}], meaning that the population of ACOs could contain a significant fraction of objects that are originally from the main asteroid belt and are not of cometary origin \cite{ref-ziffer05,ref-licandro08,ref-kim14}.  However, more detailed dynamical studies of these objects that rely on dynamical properties other than $T_J$ and instead utilize detailed numerical simulations can potentially give a better sense of objects' origins \cite{ref-fernandez14}, and near-dormant comets can also be identified simply by performing deep searches for residual activity. Such searches have in fact occasionally turned up evidence of low-level activity (e.g., in the cases of P/2003 WY$_{25}$=289P/Blanpain and (3552) Don Quixote, among others \cite{ref-jewitt06,ref-mommert14}), 
but this is not the case for most ACOs.


\subsection{Active Asteroids\label{section:aas}}

\subsubsection{Overview\label{section:aasoverview}}

In contrast to ACOs, which are observationally asteroidal and dynamically cometary (i.e., have $T_J<3$), active asteroids are observationally cometary and dynamically asteroidal (i.e., have $T_J>3$) \cite{ref-jewitt15b}.  The first type of active asteroids to come to light were the main-belt comets (MBCs), which orbit completely within the main asteroid belt yet exhibit comet-like mass loss due (at least in part) to the sublimation of volatile ices \cite{ref-hsieh06,ref-hsieh16b}.  In recent years, however, it has become increasingly clear that comet-like mass loss events due to a unexpectedly broad range of effects or combinations of effects are possible for objects on asteroid-like orbits (leading to Bauer et al.\ to suggest the more general term of active main-belt objects, or AMBOs \cite{ref-bauer12}).  In addition to sublimation, two of the more common activity generation mechanisms that have been inferred are impact disruption and rotational destabilization.  Active asteroids for which comet-like activity is not at least partially attributed to sublimation, and instead appears to be due to impacts, rotation, or other disruptive effects, are often referred to as disrupted asteroids \cite{ref-hsieh12a}.  For detailed discussions of these objects, interested readers are referred to recent reviews of MBCs\cite{ref-bertini11,ref-hsieh16b} and of active asteroids in general \cite{ref-jewitt15b}, though I will highlight key aspects of these objects here.

\subsubsection{Main-Belt Comets\label{section:mbcs}}

First recognized as a new class of cometary objects in 2006 \cite{ref-hsieh06}, MBCs have attracted interest because the unexpected present-day ice (given their location in the warm inner solar system) implied by their apparently sublimation-driven activity mean that they could potentially be used as compositional probes of inner solar system ice and also as a means of testing hypotheses that icy objects from the main-belt region of the solar system could have played a significant role in the primordial delivery of water to the terrestrial planets \cite{ref-morbidelli00,ref-raymond04,ref-obrien06,ref-hsieh14}.
While both meteoritic and spectroscopic evidence of past water in main-belt asteroids has long been available in the form of detections of hydrated minerals in both meteorites linked to asteroids in the main belt and remote spectroscopy of the asteroids themselves \cite{ref-hiroi96,ref-burbine98,ref-keil00,ref-rivkin02}, the activity observed for MBCs points to the existence of currently present water ice in the asteroid belt that could represent some of the best preserved material from this part of the solar system available for study today.  This potential for using MBCs as compositional probes of the early inner solar system is further bolstered by dynamical studies that indicate that many of them appear to be dynamically stable and are therefore unlikely to be classical comets from the outer solar system that have somehow evolved onto main-belt-like orbits \cite{ref-haghighipour09,ref-hsieh12b,ref-hsieh12c,ref-hsieh13,ref-hsieh16a}.

Beginning with the analysis of the first MBC to be discovered, 133P/Elst-Pizarro, the presence of sublimation-driven dust emission in MBCs has typically had to be inferred indirectly \cite{ref-hsieh04}.  Currently, repeated dust emission events near perihelion passages with intervening periods of quiescence is considered one of the strongest observable indicators that an object is exhibiting sublimation-driven activity \cite{ref-hsieh12a}.  This is because other potential activity drivers (e.g., impact or rotational disruption) are not expected to be able to plausibly produce such behavior at such high frequencies (elapsed times between subsequent perihelion passages, i.e., orbit periods, for MBCs tend to be on the order of 5-6 years, much shorter than the expected timescale for repeated impacts of sufficient size on a single object) or regularity (neither impacts or rotational disruptions are expected to reliably occur on timescales coinciding with an object's orbit period).

Finson-Probstein-type numerical dust modeling \cite{ref-finson68} can also be used to help determine the nature of observed comet-like dust emission from a MBC candidate by constraining the duration and other characteristics of an emission event.  This type of analysis models the dynamical behavior of dust grains of various sizes ejected at various times and aims to infer the properties of a dust emission event (e.g., particle size distribution, ejection velocities, emission duration) by determining the model parameters needed to reproduce a comet's observed activity \cite{ref-hsieh09,ref-hsieh11,ref-hsieh12b}.  The use of Finson-Probstein-type modeling to identify probable sublimation-driven activity relies on the assumption that as long as exposed ice on an object remains at a sufficient temperature to sublimate and also is not yet fully depleted, mass loss will persist in a sustained manner, meaning that model parameters corresponding to long-duration emission events should produce the best match to observations.  Meanwhile, impact disruption events are expected to produce nearly impulsive dust emission events (i.e., occurring at the time of impact) with little or no additional dust production following the initial impact event, meaning that models of impulsive or at least short-duration emission should produce the best fits to observations.



Such indirect analyses are needed to infer the action of sublimation in producing the observed activity of currently recognized MBCs (or MBC candidates) because while many attempts to confirm the presence of outgassing for a MBC using spectroscopy have been made over the last several years, none has succeeded to date.  Rather than ruling out the existence of gas emission entirely, however, these non-detections simply indicate that production rates of any gases produced from these objects are lower than the detection limits of the observations conducted to date.  These observations typically involve either searching for CN emission at 3889\AA\ (commonly used as a tracer of water sublimation in JFCs \cite{ref-bockeleemorvan04}) using large (8-10~m-class) ground-based telescopes like Keck I, Gemini North and South, the Gran Telescopio Canarias, and the Very Large Telescope facility, or searching directly for H$_2$O via its 1$_{10}$$-$1$_{01}$ ground state rotational transition at 557 GHz from space using the {\it Herschel Space Telescope}.  Upper limits to CO and CO$_2$ production rates ($Q_{\rm CO}$$\,<\,$10$^{26}$~mol~s$^{-1}$; $Q_{\rm CO_2}$$\,<\,$10$^{25}$~mol~s$^{-1}$) have also been derived from observations of active asteroids by the {\it WISE} spacecraft over the course of its survey operations \cite{ref-bauer12}.

Upper limit water production rates on the order of $Q_{\rm H_2O}$$\,\sim\,$1$\times$10$^{24}$$-$10$^{26}$~mol~s$^{-1}$ have been inferred from the aforementioned CN searches \cite[and references within]{ref-hsieh16b}, although these inferred H$_2$O production rates assume cometary CN/H$_2$O mixing ratios similar to those measured for JFCs.
Given thermal modeling results suggesting that MBCs could be depleted in HCN, the parent species of CN, relative to H$_2$O \cite{ref-prialnik09}, such conversions may not be appropriate for MBCs, meaning that upper limits on H$_2$O production rates estimated from CN searches could be too low by a large margin.
Direct searches for H$_2$O, such as those conducted for MBCs 176P and P/2012 T1 using {\it Herschel}, which yielded upper limit water production rates of $Q_{\rm H_2O}$$\,\sim\,$4$\times$10$^{25}$~mol~s$^{-1}$ and $Q_{\rm H_2O}$$\,\sim\,$7.63$\times$10$^{25}$~mol~s$^{-1}$, respectively \cite{ref-devalborro12,ref-orourke13}, may be more reliable.

Water vapor outgassing {\it has} been directly detected for the main-belt dwarf planet (1) Ceres \cite{ref-kuppers14}, but the much larger size of Ceres compared to the km-scale (or smaller) MBCs likely means that it occupies a very different physical regime and cannot be easily related to those smaller objects.  Facilities that are newly operational or will become operational in the near future, such as the Atacama Large Millimeter/submillimeter Array and the James Webb Space Telescope may provide upgraded prospects for volatile detections for MBCs \cite{ref-bockeleemorvan08,ref-kelley16}, but it is also possible that making such a detection may ultimately require an in situ investigation by a visiting spacecraft.


\subsubsection{Disrupted Asteroids\label{section:disrupted}}

The first two suspected collisionally disrupted objects in the main asteroid belt, P/2010 A2 (LINEAR) and (596) Scheila, were both first discovered to be active in 2010 \cite{ref-jewitt10b,ref-jewitt11,ref-snodgrass10,ref-bodewits11}.  Dust modeling showed fairly unequivocally that the dust emission event that took place on (596) Scheila was due to an oblique impact that caused an asymmetric ejecta cone that then evolved under the influence of radiation pressure \cite{ref-ishiguro11b}, though an additional numerical modeling analysis later suggested that P/2010 A2's dust emission might instead have been due to rotational destabilization \cite{ref-agarwal13}, though this interpretation remains controversial.  While not commonly considered before, rotational destabilization has since gained increasing acceptance as a plausible cause of observed mass shedding by asteroids, notably in cases where an asteroid may already be rotating at close to its disruption limit and is then accelerated past that limit by radiative torques (e.g., the Yarkovsky-O'Keefe-Radzievskii-Paddack, or YORP, effect) \cite{ref-rubincam00,ref-jewitt13b}.

Disruptions of main-belt asteroids present opportunities to probe the physical properties of individual asteroids as well as the population at large.  It has been shown empirically that impacts can produce observable changes in the surface of an asteroid \cite{ref-bodewits14}, presumably due to either the addition of material from the impactor to the area immediately surrounding the impact site, the excavation of subsurface material on the impacted asteroid, or both. Theoretical modeling suggests that similar observable changes could be produced as a result of mass shedding associated with rotational destabilization events \cite{ref-scheeres15}.  Thus, if various mitigating effects (e.g., the contribution of impactor material to the changed spectroscopic surface properties of collisionally disrupted asteroids) can be untangled, the study of disrupted asteroids could have significant implications for advancing our understanding of space weathering, which is the process by which exposure to micrometeorite bombardment, the solar wind, and cosmic-ray ions can change the spectral properties of the surfaces of solar system bodies such as asteroids or the Moon \cite{ref-gaffey10,ref-brunetto15}. 
Understanding space weathering essentially amounts to understanding how the properties of surface material (typically the only material that we can actually observe) on asteroids differs from those asteroids' interior material, which of course constitutes the vast majority of their bulk content.  By providing cases where both weathered surface material and exposed interior material are simultaneously present on the same asteroid, 
disrupted asteroids could serve as natural laboratories for researchers seeking to understand this important phenomenon.

The properties of collisional and rotational disruption events can also reveal information about the internal structure and material properties of the disrupted asteroids.  Analytical models coupled with information gleaned from laboratory tests can be used to predict outcomes (e.g., crater sizes) from impacts between bodies of different compositions and internal porosities and at different velocities and incidence angles, which can then be compared to observations \cite{ref-holsapple09,ref-housen03,ref-leliwakopystynski16}.  Similarly, theoretical models of asteroids with variations in their internal structure predict different failure (i.e., disruption) modes \cite{ref-hirabayashi15a,ref-hirabayashi15b}, enabling investigators to use observed rotational disruption events to constrain the material properties of the disrupted bodies \cite{ref-hirabayashi14}. 

Looking at the broader picture, knowing the rates at which impact and rotational disruptions occur and their spatial distribution in the asteroid belt could help investigators constrain the characteristics of the population of otherwise unobservable meter- or 10-meter-scale main-belt objects that may play the dominant role of impactors in impact disruptions \cite{ref-hsieh09a,ref-jewitt11} as well as provide constraints on the likelihood of rotational disruptions for different asteroids based on their material properties \cite{ref-toth06b}.  Researchers might then be able to refine evolutionary models of the asteroid belt for which the shaping of the size distribution of asteroids by collisional and rotational processes and the population of very small bodies (diameters of $<\,$1~km) provide important constraints \cite{ref-bottke05,ref-jacobson14}. 
Of course, accomplishing these objectives first relies on the continued systematic discoveries of disruption events, which is an area in which current surveys are becoming reasonably proficient, although discovery rates are still far below what is necessary to perform statistically meaningful studies anytime in the near future \cite{ref-hsieh15}.

\subsubsection{Other Active Asteroids\label{section:otheraas}}

While all MBCs (by definition) and most known disrupted asteroids orbit in the main asteroid belt, a few active objects (or suspected active objects) do not reside in the asteroid belt, yet have orbits with $T_J>3$ (implying that they at least originated in the main asteroid belt) and so therefore satisfy the definition of active asteroids.  Notable examples of such objects include asteroids (2201) Oljato and (3200) Phaethon, and Comet 107P/Wilson-Harrington (also known as (4015) Wilson-Harrington).

Oljato was associated with interplanetary magnetic field enhancements detected by the Pioneer Venus spacecraft that have been attributed to possible comet-like mass loss from the object \cite{ref-russell84}, and also once exhibited unexpectedly high ultraviolet reflectance perhaps due to cometary outgassing \cite{ref-mcfadden93}.  
No visible dust emission has ever been observed for the object, though. Meanwhile, Phaethon has been believed to be dynamically linked with the annual Geminid meteor shower for many years \cite{ref-halliday88,ref-gustafson89}, suggesting that comet-like mass loss from the object is responsible for producing the Geminid meteor stream, but like Oljato, has been observed to be visibly inactive for most of that time \cite{ref-hsieh05}. Visible dust emission was finally detected in Phaethon very recently by NASA's \textit{STEREO} solar observatory which was able to observe the object while it was extremely close to the Sun.  This dust emission has been attributed to thermal fracturing, caused by high surface temperatures experienced by the body near and during its closest approach to the Sun of 0.14 AU during perihelion \cite{ref-jewitt10a,ref-jewitt13a}.  Lastly, 107P was discovered to be active on only one occasion in 1949, exhibiting an ion tail determined to be comprised of H$_2$O and CO \cite{ref-fernandez97}, but since then, has not been seen to be active again \cite{ref-ishiguro11a}.  

Phaethon's spectrum has been reported to be similar to B-type asteroids \cite{ref-licandro07} and 107P has been classified as a C-type asteroid \cite{ref-urakawa11}, consistent with both objects originating in the outer asteroid belt, as their current dynamical properties (i.e., $T_J>3$) appear to suggest.  The spectroscopic classification of Oljato has been considerably more difficult to ascertain, but like Phaethon and 107P, its spectral properties at least appear to be more consistent with asteroids than with the nuclei of classical comets \cite[and references within]{ref-lazzarin96}.

\subsubsection{Activity Generation Mechanisms in Active Asteroids}

While this section on active asteroids has focused on sublimation, impact disruption, and rotational destabilization, there are other mechanisms that could potentially produce comet-like mass loss on asteroids, including thermal fracturing (see Phaethon, above, or the case of 322P/SOHO 1 \cite{ref-knight16}), radiation pressure sweeping, and electrostatic levitation \cite{ref-jewitt15b}.  Furthermore, it is important to consider that activity for a given active asteroid could in fact be produced by a combination of effects, and individual active episodes on the same active asteroid could even have different sets of driving mechanisms each time.

For example, 133P's activity has been observed on multiple separate occasions and is therefore strongly believed to be sublimation-driven.  However, 133P is also known to have a short rotation period ($P_{\rm rot}$=3.471~hr) and may have experienced an impact in the past that exposed buried ice, triggering its current sublimation-driven activity \cite{ref-hsieh04}.  Thus, when it was first activated, 133P's activity could have been said to be produced by a {\it combination} of sublimation, impact disruption, and rotational destabilization (where the object's rotation does not cause mass loss on its own, but instead lowers the effective gravity of the body, allowing particles launched by extremely weak sublimation to reach escape velocity and produce observable dust emission).  Then, in subsequent active episodes, 133P's continued activity could be said to be due to a combination of just sublimation and rapid rotation.
Similarly, the disintegration of P/2013 R3 (PANSTARRS) has been attributed to rotational destabilization due to the inability of gas pressure from sublimation to cause the catastrophic disintegration of the body.  However, while rapid rotation may have caused the breakup of the body, dust production was observed to continue over several months for the individual fragments, suggesting the additional contribution of sublimation, perhaps of icy interior material exposed by the fragmentation event, to the object's ongoing activity \cite{ref-jewitt14a}.

Unfortunately, it is not always possible to confirm or rule out all of the possible causes of observed activity in active asteroids.  Many active asteroid nuclei are extremely small (diameters of $\ll\,$1~km), requiring large ($>\,$8~m) telescopes to detect while they are distant and inactive (which is the only time when nucleus properties can be measured without dust contamination) \cite{ref-maclennan12,ref-hsieh14b}, if they can be detected at all.  In these cases, high-quality lightcurves to determine rotation rates are extremely difficult to obtain, complicating efforts to determine the plausibility of rotational disruption scenarios.  As mentioned above, recurrent activity and dust modeling can be used to infer the action of sublimation, but confirming the presence of recurrent activity is a time-consuming process (typically requiring waiting for $\sim$5-6~years after an active asteroid is first discovered for it to return to perihelion, assuming favorable observing conditions during that next perihelion passage, which is not always the case), while dust models can be frequently underconstrained by observations making unambiguous conclusions difficult to reach at times.

Distinguishing different types of active asteroids is a topic of great importance in this research area as it directly impacts the strategies used to study individual objects as well as the interpretation of the abundance and distribution of these objects as more are discovered.  As such, in the future, the development of new observational, analytical, and theoretical techniques for ascertaining the sources of observed activity in active asteroids, particularly for inferring or even directly detecting sublimation, will be extremely valuable.

\subsection{Other Continuum Objects}

There are many other types of continuum objects for which there is not enough space in this brief review to discuss in detail, but which are worth at least mentioning.  For instance, cometary activity has long been linked to the production of debris streams which cause regular meteor showers on Earth \cite{ref-jenniskens02,ref-wiegert05}.  While there is some uncertainty associated with linking known solar system bodies to meteor streams given the possibility of chance orbital alignments \cite{ref-wiegert04}, the association of a meteor stream with an apparently inactive object is still highly suggestive that that object is at least potentially cometary \cite{ref-jenniskens08}.  Many are ACOs (Section~\ref{section:continuumobjs}\ref{section:acos}), suggesting they really are dormant comets, but others are not \cite{ref-olssonsteel88}.  Dynamically asteroidal ($T_J>3$) meteor stream parents are interesting because they either have asteroidal origins (i.e., are from the main belt), raising questions as to the causes of their mass loss, or are classical comets (i.e., from the outer solar system) that have evolved onto asteroid-like orbits, raising questions as to how often this occurs and the impact on our estimates of the cometary contribution to the NEO population.  The best known dynamically asteroidal meteor parent body is the Geminid parent, (3200) Phaethon (Section~\ref{section:continuumobjs}\ref{section:aas}\ref{section:otheraas}), though (1566) Icarus, (2101) Adonis, (2201) Oljato (Section~\ref{section:continuumobjs}\ref{section:aas}\ref{section:otheraas}), and (2212) Hephaistos, among others, all with $T_J>3$, have also been associated with meteor streams \cite{ref-olssonsteel88}.

The Hilda asteroids, found beyond the main asteroid belt in the 3:2 mean-motion resonance with Jupiter at $\sim$4.0~AU, and the Jovian Trojans, found at Jupiter's L4 and L5 Lagrange points, are also notable.  Both groups are dominated by primitive asteroid types (C, P, and D) \cite{ref-dahlgren95,ref-dahlgren97,ref-demeo13,ref-bendjoya04}, where D-type asteroids are spectroscopically similar to comet nuclei.  As such, although the Kuiper Belt and scattered disk are the commonly accepted primary sources of JFCs \cite{ref-volk08}, Hildas and Trojans are thought to perhaps also contribute to the JFC population \cite{ref-disisto05,ref-toth06a,ref-marzari95,ref-guilbertlepoutre14}.  Meanwhile, Centaurs (some of which exhibit activity) are widely viewed as the dynamical transition state between Kuiper Belt and scattered disk objects, and JFCs \cite{ref-levison97}.

Recently, low-activity comets on LPC-like orbits, or so-called ``Manx comets'', have gained attention as possible probes of early planetary migration models.  Related to the class of solar system objects known as Damocloids, which are inactive objects with $T_J\leq2$ and may be inactive HTCs and LPCs \cite{ref-jewitt05}, Manx comets have been hypothesized to be objects from the inner solar system, some of which might be made of distinctly un-comet-like, mostly non-icy material, that were ejected into the Oort cloud during the era of planetary migration but have recently been perturbed onto orbits bringing them back into the inner solar system \cite{ref-meech16}.  By studying their properties and determining the size of the steady-state population, researchers hope to constrain the quantity of inner solar system material that was originally implanted into the Oort cloud due to large scale planetary migration early in the solar system's history, which in turn will inform assessments of the plausibility of competing solar system evolution models.







\section{Dynamics\label{section:dynamics}}

I wish to conclude this review with a brief discussion of various dynamical issues related to asteroids and comets.  Traditionally, the Kuiper belt has been regarded as the source of the JFCs while the Oort cloud has been regarded as the source of HTCs and LPCs \cite{ref-weissman99,ref-volk08,ref-wang14}. 
As discussed above (Section~\ref{section:continuumobjs}\ref{section:acos}), the use of $T_J$ also provides a simple but generally effective means for determining whether an object possesses a ``comet-like'' orbit ($T_J<3$) or an ``asteroid-like'' orbit ($T_J>3$).  Recent dynamical analyses as well as the discovery of the MBCs, which demonstrate that main-belt objects can and do contain sufficient volatile material to drive present-day cometary activity, however, have complicated these neat dynamical principles.

Studies have been performed over the years to investigate whether JFCs could transition onto asteroidal ($T_J>3$) orbits, with recent results indicating that this is indeed possible \cite{ref-fernandez02,ref-levison06}.  The formulation of $T_J$ is imperfect as it relies on an idealized approximation of the solar system in which the Sun and Jupiter are the only massive objects and Jupiter's orbit is perfectly circular and non-inclined.  Missing from that formulation are consideration of Jupiter's real-world non-zero eccentricity and inclination, the gravitational effects of the other planets (which admittedly typically are not particularly significant, except in the cases of extremely close encounters), and the effect of non-gravitational forces such as the Yarkovsky effect and cometary outgassing \cite{ref-pittich04}.

A recent dynamical study consisting of numerical integrations of a large number of synthetic test particles and all of the major planets and the Sun as gravitational perturbers found that even without the consideration of non-gravitational forces, $T_J$ is not conserved for a non-negligible fraction of originally ``comet-like'' objects as well as for a significant fraction of originally ``asteroid-like'' objects \cite{ref-hsieh16a}.  This study found that test particles with initially JFC-like orbits were able to evolve onto main-belt-like orbits with $T_J>3$ on relatively short timescales ($<\,$2~Myr) and test particles with initially main-belt-like orbits were able to evolve onto JFC-like orbits on similar timescales. Intriguingly, if these main-belt objects were able to maintain significant reservoirs of preserved ice until they reach such orbits, it is not inconceivable that they could then become indistinguishable from ``ordinary'' JFCs from the outer solar system, raising the intriguing possibility that studies to date of JFCs may have included at least a few objects that were not from the outer solar system at all, and instead actually hail from the main asteroid belt.

In the aforementioned integrations, the effects of repeated close encounters with the terrestrial planets appeared to be the major drivers of initially JFC-like particles onto main-belt-like orbits.  Fortunately for researchers with hopes of using MBCs as tracers of ice in the asteroid belt, which needless to say, would be greatly complicated by significant JFC contamination of the main belt, the JFC interlopers in these integrations were seen to stay confined to high-eccentricity, high-inclination orbits, and did not reach the low-eccentricity, low-inclination orbits occupied by many, but not all, of the MBCs.  This is consistent with previous dynamical simulations that found similar behavior for a number of dynamical clones of real JFCs \cite{ref-fernandez02}.

The possibility that some apparent members of the JFC population could in fact be from the main asteroid belt is further supported by work showing that a handful of JFCs appear to have substantially more stable orbits than the average JFC, and that several of those objects also show unusually low activity relative to other comets \cite{ref-fernandez15}.  The observational studies discussed in Section~\ref{section:continuumobjs}\ref{section:acos} also found several objects with high albedos and un-comet-like spectroscopic taxonomic types (e.g., S or Q) among the ACO population, inconsistent with JFC origins and more consistent with origins in the main asteroid belt \cite{ref-fernandez05,ref-demeo08,ref-kim14,ref-binzel04,ref-licandro06}.

It should be noted that all of the issues discussed above assume the modern configuration of the solar system.  If significant planetary migration occurred early in the solar system's history as various dynamical models suggest, even more mixing of inner and outer solar system material (e.g., implantation of trans-Neptunian objects in the asteroid belt or of asteroidal material in the Oort Cloud \cite{ref-levison09,ref-meech16}) may have occurred in the past, further emphasizing the importance of a coherent observational and dynamical picture of the solar system that considers the full range of continuum objects for achieving a complete understanding of our origin and history.





\section{Conclusions}

In summary, the population of small bodies in our solar system today, including both minor planets and classical comets, is far less well-delineated into distinct groups of objects than the classical paradigm might have led one to believe in the past.  These objects instead appear to occupy a continuum spanning the full range of observational, physical, and dynamical properties classically attributed solely either to asteroids or comets.  We now know of currently actively sublimating main-belt objects that could have originated either in the asteroid belt or in the outer solar system, and objects displaying comet-like activity that may have no volatile ice content whatsoever.  We have found objects composed of inner-asteroid-belt-like material on long-period comet-like orbits, and active objects on comet-like orbits that may in fact originate from the asteroid belt. We also now recognize that dormant comets may be found both comet-like and un-comet-like orbits.  The population of continuum objects is extraordinarily diverse, with each type of object holding the potential for revealing exciting new insights about our solar system due to their unique sets of overlapping comet-like and asteroid-like properties.
Given this complexity and the growing interest in addressing the many questions that it has raised thus far, it is likely that many more interesting findings await us in this rapidly developing field in the coming years.

\enlargethispage{20pt}




\competing{The author declares that he has no competing interests.}

\funding{HHH gratefully acknowledges travel funding provided by the Royal Society, and research support provided by the NASA Planetary Astronomy program under grant NNX14AJ38G and the NASA Solar System Observations program under grant NNX16AD68G.}




\end{document}